\documentclass[aps,prl,twocolumn,amsmath,showpacs,letterpaper,floatfix]{revtex4}

\usepackage{graphicx,amsmath,amssymb}
\usepackage{color}

\newcommand{\eac}[0]{{\it et al.}, }

\begin{document}

\title{Memory for Light as a Quantum Process}

\author{M. Lobino,$^1$ C. Kupchak,$^1$ E. Figueroa,$^{1,2}$ and A. I. Lvovsky$^{1,}$}
\email{lvov@ucalgary.ca}

\affiliation{$^1$ Institute for Quantum Information Science,
University of Calgary, Calgary, Alberta T2N 1N4, Canada}

\affiliation{$^2$ Max-Planck-Institut f{\"u}r Quantenoptik,
Hans-Kopfermann-Str.\ 1, 85748 Garching, Germany}

\begin{abstract}
We report complete characterization of an optical memory based on
electromagnetically induced transparency. We recover the
superoperator associated with the memory, under two different working
conditions, by means of a quantum process tomography technique that
involves storage of coherent states and their characterization upon
retrieval. In this way, we can predict the quantum state retrieved
from the memory for any input, for example, the squeezed vacuum or the Fock state. We
employ the acquired superoperator to verify the nonclassicality
benchmark for the storage of a Gaussian distributed set of coherent
states.
\end{abstract}

\pacs{42.50.Ex, 03.67.-a, 32.80.Qk, 42.50.Dv}
 \maketitle

\paragraph{Introduction}
Quantum memory for light is an essential technology for long
distance quantum communication \cite{DLCZ} and for any future
optical quantum information processor. Recently, several experiments
have shown the possibility to store and retrieve nonclassical states
of light such as the single photon
\cite{KuzmichSingle,LukinSingle}, entangled \cite{KimbleEntang} and squeezed vacuum
\cite{KozumaStorage,LvovskyStorage} states using coherent interactions with
an atomic ensemble.

In order to evaluate the applicability of a quantum memory apparatus for practical quantum communication and computation, it is insufficient to know its performance for specific, however complex, optical states, because in different protocols, different optical states are used for encoding quantum information \cite{DLCZ,Lloyd}. Practical applications of memory require answering a more
general question: how
will an \emph{arbitrary} quantum state of light be preserved after
storage in a memory apparatus?

Here we answer this question by performing complete characterization
of the quantum process associated with optical memory based on
electromagnetic induced transparency  (EIT)
\cite{FleischhauerReview}. Memory characterization is achieved by
storing coherent states (i.~e. highly attenuated laser pulses) of
different amplitudes and subsequently measuring the quantum states
of the retrieved pulses. Based on the acquired information, the
retrieved state for any arbitrary input can be predicted and
additionally, any theoretical benchmark on quantum memory
performance can be readily verified.

\paragraph{Coherent state quantum process tomography} We can define complete characterization of an optical quantum memory
as the ability to predict the retrieved quantum state $\hat{\mathcal{E}}(\hat\rho)$ when
the stored input state $\hat\rho$ is known. This is a particular case of
the quantum ``black box" problem, which is approached through a
procedure called quantum process tomography (QPT) \cite{MohseniQPT}.
\begin{figure}[b]
  \center{\includegraphics{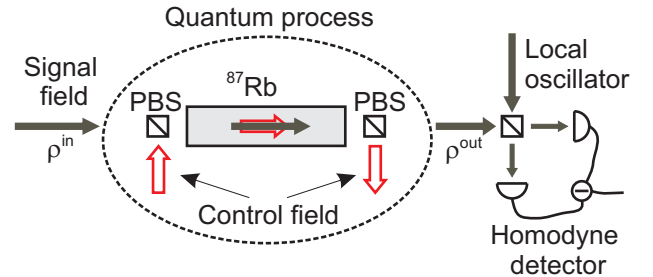}}
  \caption{(color online). Schematic of the experimental setup used
  to characterize the process associated with the quantum memory.
  PBS, polarizing beam splitter.}
  \label{fig.1}
\end{figure}
QPT is based on the fact that every quantum process (in our case,
optical memory) is a linear map on the linear space $\mathbb{L}(\mathbb{H})$ of
density matrices over the Hilbert space $\mathbb{H}$ on which the
process is defined. The associated process can thus be characterized
by constructing a spanning set of ``probe" states in $\mathbb{L}(\mathbb{H})$ and
subjecting each of them to the action of the quantum
``black box". If we measure the process output $\hat{\mathcal{E}}(\hat\rho_i)$ for each
member $\hat\rho_i$ of this spanning set, we can calculate the process
output for any other state $\hat\rho=\sum_i a_i\hat\rho_i$ according to
\begin{equation}
    \hat{\mathcal{E}}(\hat\rho)=\sum_i a_i\hat{\mathcal{E}}(\hat\rho_i).
    \label{Eq.linearity}
\end{equation}

The challenge associated with this approach is the construction of
the appropriate spanning set, given the infinite dimension of the
optical Hilbert space and the lack of techniques for universal
optical state preparation. For this reason, characterizing memory
for light, that is not limited to the qubit subspace, is much more
difficult than memory for superconducting qubits, which has been
reported recently \cite{NeeleyQPT}. Our group has recently developed
a process characterization technique that overcomes these challenges
\cite{LobinoQPT}. Any density matrix $\hat\rho$ of a quantum optical
state can be written as a linear combination of density matrices of
coherent states $|\alpha\rangle$ according to the optical
equivalence theorem
\begin{equation}
    \hat\rho=2\int P_{\hat\rho}(\alpha)|\alpha\rangle\langle\alpha|d^2\alpha,
    \label{Eq.Pfunction}
\end{equation}
where $P_{\hat\rho}(\alpha)$ is the state's Glauber-Sudarshan P-function
and the integration is performed over the entire complex plane.
Although the P-function is generally highly singular, any quantum
state can be arbitrarily well approximated by a state with an
infinitely smooth, rapidly decreasing P-function \cite{Klauder}.
Therefore, by measuring how the process affects coherent states, one
can predict its effect on any other state. The advantage of such
approach (which we call coherent-state quantum process tomography or
csQPT) is that it permits complete process reconstruction using a
set of ``probe" states that are readily available from a laser.

\paragraph{Experimental setup}
We performed csQPT on optical memory \cite{LvovskyStorage} realized in
a warm rubidium vapor  by means of electromagnetically-induced
transparency (Fig.~\ref{fig.1}). The atoms are $^{87}$Rb and the vapor
temperature is kept constant at 65$^\circ$C.

The signal field is resonant with the $|^5 S_{1/2}, F=1
\rangle\leftrightarrow|^5 P_{1/2}, F=1\rangle$ transition at 795 nm
and is produced by a continuous-wave Ti:Sapphire laser. An external
cavity diode laser, phase locked at 6834.68 MHz to the signal laser
\cite{AppelPhaselock} serves as the EIT control field source, and is
resonant with the $|^5 S_{1/2}, F=2 \rangle\leftrightarrow|^5
P_{1/2}, F=1\rangle$ transition. The fields are red detuned from
resonance by 630 MHz in order to improve the storage efficiency.
The control field power is 5 mW and
the beam spatial profile is mode matched with the signal beam to a
waist of 0.6 mm inside the rubidium cell. Signal and control fields are
orthogonally polarized; they are mixed and separated using polarizing beam splitters.

The two photon detuning $\Delta_2$ between the signal
and control fields is modified by varying the
frequency of the control field laser through the phase lock circuit, while an acousto-optical modulator (AOM)
is used to switch on and off the control field intensity.
We analyzed two different operative conditions
characterized by $\Delta_2$ = 0 and 0.54 MHz.

\begin{figure}
  \center{\includegraphics{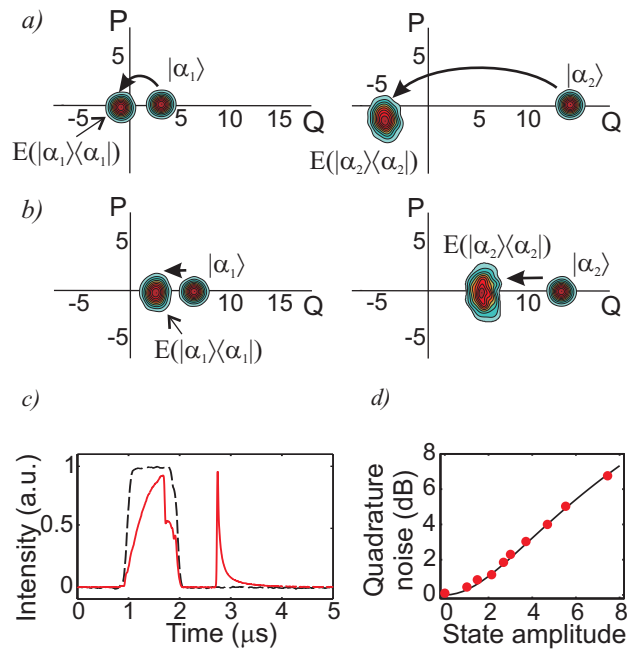}}
  \caption{(color online). Wigner functions of input
  coherent states with two different amplitudes and
  the corresponding retrieved states
  $\hat{\mathcal{E}}(|\alpha_1\rangle\langle\alpha_1|)$ and
  $\hat{\mathcal{E}}(|\alpha_2\rangle\langle\alpha_2|)$. The input state amplitudes are
  $\alpha_1=2.3$,  $\alpha_2=10.3$ (a) and  $\alpha_1=4.5$,  $\alpha_2=10.9$ (b).
  Two-photon detunings between the control and signal fields are 0.54 MHz (a) and 0 MHz (b).
  Input pulse (black dashed line) and retrieved light (red solid line) (c).
  Variance of the phase quadrature as a function of the retrieved state amplitude (d).
  }
  \label{fig.2}
\end{figure}

The input pulse is
obtained by chopping the continuous-wave signal beam via an AOM to
produce 1 $\mu$s pulses [Fig.~\ref{fig.2}(c)] with a 100 kHz
repetition rate. A second AOM is used to compensate for the
frequency shift generated by the first.  Transfer of the light state
into the atomic ground state superposition (atomic spin wave) is
accomplished by switching the control field off for the storage duration of $\tau=1\ \mu$s when the input pulse
is inside the rubidium cell.

We performed full state reconstruction of both the input and
retrieved fields by time domain homodyne tomography
\cite{LvovskyReview}. A part of the Ti:Sapphire laser beam serves
as a local oscillator for homodyne detection; while its phase is
scanned via a piezoelectric transducer, the homodyne current is
recorded with an oscilloscope. For every state, 50000 samples of
phase and quadrature are measured and processed by the maximum
likelihood algorithm \cite{LvovskyMaxLik,LvovskyMaxLik2}, estimating
the state density matrix in the Fock basis.

%

\paragraph{Tomography of quantum memory}
In order to determine the coherent state mapping necessary for
reconstructing the process, we measured 10 different coherent states
$|\alpha_i\rangle$ with mean photon numbers ranging from 0 to 285
along with their corresponding retrieved states
$\hat{\mathcal{E}}(|\alpha_i\rangle\langle\alpha_i|)$ [Fig.~\ref{fig.2}(a) and (b)].
Subsequently, we applied polynomial interpolation to determine the
value of $\hat{\mathcal{E}}(|\alpha\rangle\langle\alpha|)$ for any value of $\alpha$
in the range 0 to 16.9. Performing tomographic reconstruction for
these highly displaced states requires good phase stability between
the signal and local oscillator. Phase fluctuations produce an
artefact in the reconstruction in the form of amplitude dependent
increase in the phase quadrature variance. In our measurements, the reconstructed input
states $| \alpha_i \rangle$ resemble theoretical coherent states
with a fidelity higher than $0.999$ for mean photon values up to 150
[Fig.~\ref{fig.2}(a) and (b)].

By inspecting the Wigner functions of the input and retrieved
states, one can clearly notice the detrimental effects of the
memory. First, there is attenuation of the amplitude by a factor of
$0.41\pm0.01$ for the signal field in two-photon resonance with the
control, which increases to a factor of $0.33\pm0.02$ when a
two-photon detuning of  $\Delta_2$ = 0.54 MHz is introduced. This
corresponds to a mean photon number attenuation by factors of
$0.17\pm0.02$ and $0.09\pm0.01$, respectively. Note that in the case
of nonzero two-photon detuning, the attenuation is greater than the
factor of 0.14 obtained in classical intensity measurement [Fig.~
\ref{fig.2}(c)]. This is because the temporal mode of the retrieved
state is slightly chirped, and could not be perfectly matched to the
mode of the local oscillator.

\begin{figure}[t]
  \center{\includegraphics{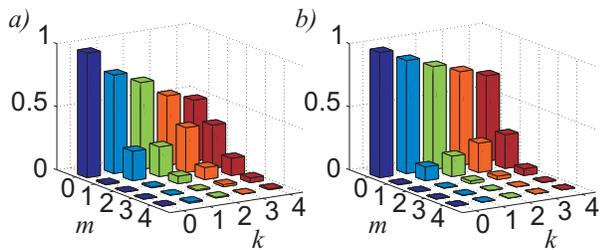}}
  \caption{(color online).
  The diagonal elements of the process tensor $\chi_{kk}^{mm}$,
  measured by csQPT in the Fock basis for $\Delta_2$ = 0 (a) and 0.54 MHz (b).}
  \label{fig.1n}
\end{figure}

Second, retrieved coherent states
experience an increase in the phase quadrature variance that depends
quadratically on the state amplitude. This effect produces an
ellipticity in the retrieved state Wigner function (Fig.~
\ref{fig.2}(a) and (b)] and can be attributed to the noise in the phase
lock between the signal and control lasers \cite{AppelPhaselock}.
Fluctuations $\Delta\phi$ of the relative phases between the two
interacting fields randomize the phase of the retrieved signal field
with respect to the local oscillator. Assuming a Gaussian
distribution for $\Delta\phi$ with zero mean and variance
$\sigma^2_\phi$ the variance of the phase quadrature can be
expressed as:
\begin{equation}
    \sigma^2_q=\frac{1}{2}+\frac{q_0^2}{2}\left(1-e^{2\sigma^2_\phi}\right),
    \label{Eq.variance}
\end{equation}
where $q_0$ is the mean amplitude. We fit our experimental data with
Eq.\ref{Eq.variance} and estimate an $11^\circ$ standard deviation
for $\Delta\phi$ [Fig.~\ref{fig.2}(d)], in agreement with independent
estimates \cite{AppelPhaselock}.

The third detrimental effect
preventing the atomic ensemble from behaving as a perfect memory is
the population exchange between atomic ground states
\cite{HetetMemory,FigueroaSlowlight}.
Besides limiting the memory lifetime, this exchange generates
spontaneously emitted photons in the signal field mode adding an
extra noise that thermalizes the stored light by increasing the
quadrature variance independently of the input amplitude and phase.
We measured the extra noise from the quadrature variance of
retrieved vacuum states $\hat{\mathcal{E}}(|0\rangle\langle 0|)$ and found it to
equal 0.185 dB when both fields were tuned exactly at the two photon
resonance, which corresponds to the mean photon number in the
retrieved mode equal to $\overline{n}=Tr\left[\hat{n}
\hat{\mathcal{E}}(|0\rangle\langle 0|) \right]=0.022$. This noise is reduced to 0.05
dB (corresponding to $\overline{n}=0.005$ ) in the presence of two
photon detuning. For this reason, it is beneficial to implement
storage of squeezed light in the presence of two-photon detuning, in
spite of higher losses.

\begin{figure}[b]
  \center{\includegraphics{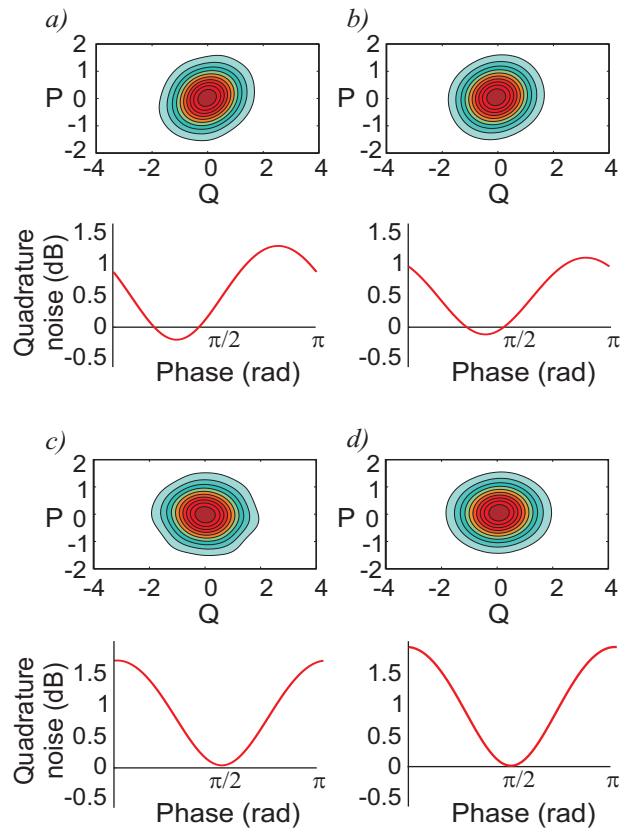}}
  \caption{(color online). Comparison of the experimentally
  measured squeezed vacuum states retrieved from the quantum
  memory and those predicted with csQPT. For each case, the
  Wigner function and the quadrature variance as a function of
  the local oscillator phase are shown. (a), Experimental
  measurement \cite{LvovskyStorage} with  $\Delta_2$ = 0.54 MHz. (b), Prediction with  $\Delta_2$ = 0.54 MHz.
  (c), Experimental measurement with $\Delta_2$ = 0. (d), Prediction with  $\Delta_2$ = 0.
  }
  \label{fig.3}
\end{figure}

In the presence of the two-photon detuning,
the evolution of the atomic ground state superposition brings about
a phase shift of the retrieved state with respect to the input by
$2\pi\Delta_2\tau$ = $200^\circ$ as is visible in Fig.~\ref{fig.2}(a).

Based on the information collected from the storage of coherent
states, we reconstruct the memory process in the $\chi$-matrix
representation, defined by \cite{NielsenBook,Chuang}
\begin{equation}
    \hat{\mathcal{E}}(\hat\rho)=\sum_{k,l,m,n}\chi^{n,m}_{k,l}A_{l,n}\hat\rho A_{m,k},
    \label{Eq.chi}
\end{equation}
where $\chi^{n,m}_{k,l}$ is the rank 4 tensor comprising full
information about the process and ${A_{i,j}}$ is a set of operators
that form a basis in the space of operators on $\mathbb{H}$. Since
$\mathbb{H}$ is the Hilbert space associated with an electromagnetic
oscillator, it is convenient to choose $A_{i,j}=|i\rangle\langle j|$,
where $|i\rangle$ and $|j\rangle$ are the photon number states. The
details of calculating the process tensor are described elsewhere
\cite{LobinoQPT}; Fig.~\ref{fig.1n} displays the
diagonal subset $\chi^{m,m}_{k,k}$ of the process tensor elements.

\paragraph{Performance tests} In order to verify
the accuracy of our process reconstruction, we have used it to
calculate the effect of storage on squeezed vacuum with $\Delta_2$ =
0.54 MHz, as studied in a recent experiment of our group \cite{LvovskyStorage},
and with  $\Delta_2$ = 0 MHz.
We applied the superoperator tensor
measured with csQPT to the squeezed vacuum produced by a
subthreshold optical parametric amplifier with
a noise reduction in the squeezed quadrature of $-1.86$ dB and noise
amplification in the orthogonal quadrature of $5.38$ dB 
(i.e. the same state as used as the memory input in Ref.~\cite{LvovskyStorage}). In this way,
we obtained a prediction for the state retrieved from the memory,
which we then compared with the results of direct experiments.
This comparison yields quantum mechanical fidelities of
$0.9959\pm0.0002$ and $0.9929\pm0.0002$ for the two-photon detunings
of $\Delta_2$ = 0.54 MHz and $\Delta_2$ = 0 respectively
(Fig.~\ref{fig.3}).

As discussed above, zero detuning warrants lower losses (thus higher
amplitude of the noise variance) and no phase rotation, but higher
excess noise (thus no squeezing in the retrieved state).
Nevertheless the two photon resonant configuration offers a better
fidelity if the single photon state is stored
\cite{KuzmichSingle,LukinSingle}, as evidenced by comparing the
superoperator element $\chi_{1,1}^{1,1}$ of Fig. \ref{fig.1n} (a) and
(b).

\begin{figure}[t]
  \center{\includegraphics{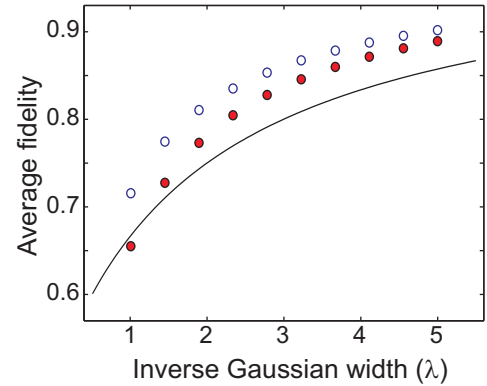}}
  \caption{(color online). Average fidelity of the
  quantum memory for a Gaussian distributed set of coherent states. Blue empty (red filled) dots show
  the average fidelity calculated from the csQPT experimental
  data for
  $\Delta_2$ = 0 (0.54 MHz). The experimental uncertainty is 0.0002.
  The solid line shows the classical limit \cite{PolzikBenchmark}.
  }
  \label{fig.4}
\end{figure}

In addition to the ability to predict the output of the memory
for any input state, our procedure can be used to estimate the
performance of the memory according to any available benchmark. As
an example, we analyze the performance of our memory with respect to
the classical limit on average fidelity associated with the storage
of coherent states with amplitudes distributed in phase space
according to a Gaussian function of width $1/\lambda$
\cite{PolzikBenchmark}. This limit as a function of $\lambda$ is given by:
\begin{equation}
    F(\lambda)=2\lambda \int_0^{+\infty}\exp{(-\lambda\alpha^2)\langle\alpha|\hat{\mathcal{E}}(|\alpha\rangle\langle\alpha|)|\alpha\rangle}\alpha
    d\alpha\leq\frac{1+\lambda}{2+\lambda}.
    \label{Eq.benchmark}
\end{equation}
From csQPT data, we evaluate the average fidelity associated with our memory for both
values of  $\Delta_2$ (Fig.~\ref{fig.4}). Both configurations show
nonclassical behavior. The higher value of average fidelity
correspond to $\Delta_2$ = 0 and is explained by a higher storage
efficiency.

\paragraph{Conclusion}
In summary, we have demonstrated complete characterization of an
EIT-based quantum memory by csQPT. This procedure allows one to
predict the effect of the memory on an arbitrary quantum-optical
state, and thus provides the ``specification sheet" of
quantum-memory devices for future applications in quantum
information technology. Furthermore, our results offer
insights into the detrimental effects that affect the storage
performance and provide important feedback for the device
optimization. We anticipate this procedure to become standard in
evaluating the suitability of a memory apparatus for practical
quantum telecommunication networks.

\paragraph*{Acknowledgements}
This work was supported by NSERC, iCORE, CFI, AIF, Quantum$Works$, iCORE (C.K.)
and CIFAR (A.L.).

\end{document}